\documentclass{PoS}
\usepackage{amsmath,amssymb}
\newcommand{\vrt}{V_T(r)}
\newcommand{\vre}{V^{\rm re}_T(r)}
\newcommand{\vim}{V^{\rm im}_T(r)}
\newcommand{\wet}{w_{\scriptscriptstyle E}(r,\tau)}    
\title{Effective thermal potential between static $Q$ and $\bar{Q}$
  in SU(3) gauge theory}

\ShortTitle{Non-perturbative thermal potential}

\author{\speaker{Dibyendu Bala}\\
        Tata Institute of Fundamental Research\\
        E-mail: \email{dibyendu@theory.tifr.res.in}}

\author{Saumen Datta\\
        Tata Institute of Fundamental Research\\
        E-mail: \email{saumen@theory.tifr.res.in}}

\abstract{A non-perturbative calculation of the effective thermal
  potential between heavy $Q$ and $\bar{Q}$ from lattice QCD is
  difficult, and usually involves a Bayesian analysis.  Here we
  present a simple method to obtain the potential from smeared Wilson
  loop, using the structure of the thermal Wilson loop. We present
  results for the $Q \bar{Q}$ thermal potential in a gluonic plasma for
  temperatures $\lesssim 2 T_c$. We also present
  preliminary results for the effective potential when the $Q$ and
  $\bar{Q}$ are in octet color configuration.}

\FullConference{37th International Symposium on Lattice Field Theory -
  Lattice2019\\
  16-22 June 2019\\
  Wuhan, China.}

\begin{document}
\section{Introduction}
Quarkonium is an important probe of quark-gluon plasma formation in
relativistic heavy ion collision experiments. One can define an
effective ``thermal potential'' to understand the medium modification
of quarkonia \cite{laine}. The thermal potential plays an important
role in the treatment of in-medium quarkonia as open quantum system
\cite{akamatsu}.  To define the potential at finite temperature
consider the real time $Q\bar Q$ operator, $M(r=|\vec x - \vec y| ,t)
= \bar \psi(\vec x,t) U(\vec x,\vec y;t) \psi(\vec y,t)$.  In the
static limit, the forward correlation function of this operator $
c(r,t) = \langle M(r,t) \, M^\dagger(r,0) \rangle $ becomes
proportional to a rectangular real time Wilson loop $w(r,t)$. The
static potential is defined as \cite{laine}
\begin{equation}
  V(r) \ = \ i \lim\limits_{t \to \infty} \frac{\partial \log
    \langle w(r,t) \rangle }
  {\partial t}.
  \label{eq.pot}
\end{equation}
Eqn. (\ref{eq.pot}) has been calculated in leading order HTL
perturbation theory. It gives a complex potential: $V(r) \; = \; \vre
\, - \, i \, \vim$, with $\vre$ the usual Debye-screened Coulomb
potential and
\begin{equation}
  \vim \ = \ \frac{2 g^2 \, T}{3 \pi} \; \int_0^\infty \, dz \;
  \frac{z}{(z^2+1)^2} \; \left( 1 - \frac{\sin z \, m_{\scriptscriptstyle D} \, r}
       {z \, m_{\scriptscriptstyle D} \, r } \right).
\label{eq.vim}\end{equation}  
Here $T$ denotes the temperature, and $m_{\scriptscriptstyle D}$ is the Debye
mass.
  
On the lattice we can only calculate Euclidean-time Wilson loop
$\wet$, defined in the time interval $\tau=[0,\beta=1/T)$. This can be
connected to the real-time Wilson loop $w(r,t)$ through a spectral
function $\rho(\omega)$ \cite{hatsuda},
\begin{equation}
  \langle \wet \rangle =\int_{-\infty}^{\infty} d\omega \ \rho(\omega)
  \ e^{-\omega \tau}.
\end{equation}
$w(r,t)$ can be constructed by taking Fourier transform of
$\rho(\omega)$. On lattice obtaining $\rho(\omega)$ is very difficult
problem as we have only small number of data points along the temporal
direction.

There have been attempts to calculate $\rho(\omega)$ using various
Bayesian analysis methods. Maximum Entropy method was used in
Ref. \cite{hatsuda}; however the potential obtained there was not
screened above $T_{c}$. Also the quality of signal for Wilson loop
deteriorates very fast with the size of the loop. To improve the
signal, we have used multilevel algorithm, slicing the lattice in
$\tau$ direction, and used APE smeared spatial links in the
construction of the Wilson loop.  If a suitable potential can be
defined, it should not depend on the smearing of the spatial link; we
have checked this by using various levels of smearing.  In the
literature Coulomb gauge fixed Wilson line correlators have often been
used to extract the potential.  Bayesian reconstruction of such a
correlator has shown a screened potential, and the imaginary part of
the potential has also been obtained \cite{rothkopf}. However the
results have large errors \cite{rothkopf}. There are also studies
using the method of moments and making some ansatz for the form
of $\rho(\omega)$ \cite{peter}.

We use a different analysis method, using the structure of the Wilson
loop; our method will be described in the next section.  In
Sec. \ref{sec.singlet} we will show the results for the potential for
gluonic plasma. More details regarding these can be found in
Ref. \cite{our}.

For QGP phenomenology, one needs the thermal potential not only between
$Q \bar{Q}$ in the singlet channel, but also when they are in an octet
color configuration. The octet potential is also a necessary ingredient
for the open quantum system approach \cite{akamatsu}. In Sec. \ref{sec.octet}
we discuss the interaction potential between the static $Q$ and $\bar{Q}$
in a color-octet configuration.

\section{Method}
\label{sec.method}
At zero temperature $ \langle w_{E}(r,\tau;T=0) \rangle \sim e^{
  -V(r)\tau }$ at large $\tau$; as a result one would see a plateau in
the effective mass $m_{eff}(r,\tau_{i})=log\frac{\langle
  w_{E}(r,\tau_{i}) \rangle}{\langle w_{E}(r,\tau_{i}+1)\rangle}$.
$V(r)$ determined from lattice is qualitatively similar to the Cornell
potential.  However at finite temperature above $T_c$ there is no
plateau in the effective mass. Given that $\vrt$ is expected to have
an imaginary part \cite{laine}, one should not expect a plateau either.

Motivated from HTL perturbation theory, we split $\wet$ as:
\begin{equation}
  \log \, (\langle w_{E}(r,\tau) \rangle) \ = \ \frac{1}{2} \, \log
  \left(\frac{\langle w_{E}(r,\tau)\rangle}{\langle w_{E}(r,\beta-\tau)
  \rangle}\right) \; + \; \frac{1}{2} \log \, ({\langle w_{E}(r,\tau)\rangle}
       {\langle w_{E}(r,\beta-\tau)\rangle}).
\end{equation}

\begin{figure}
\centerline{\includegraphics[width=7cm]{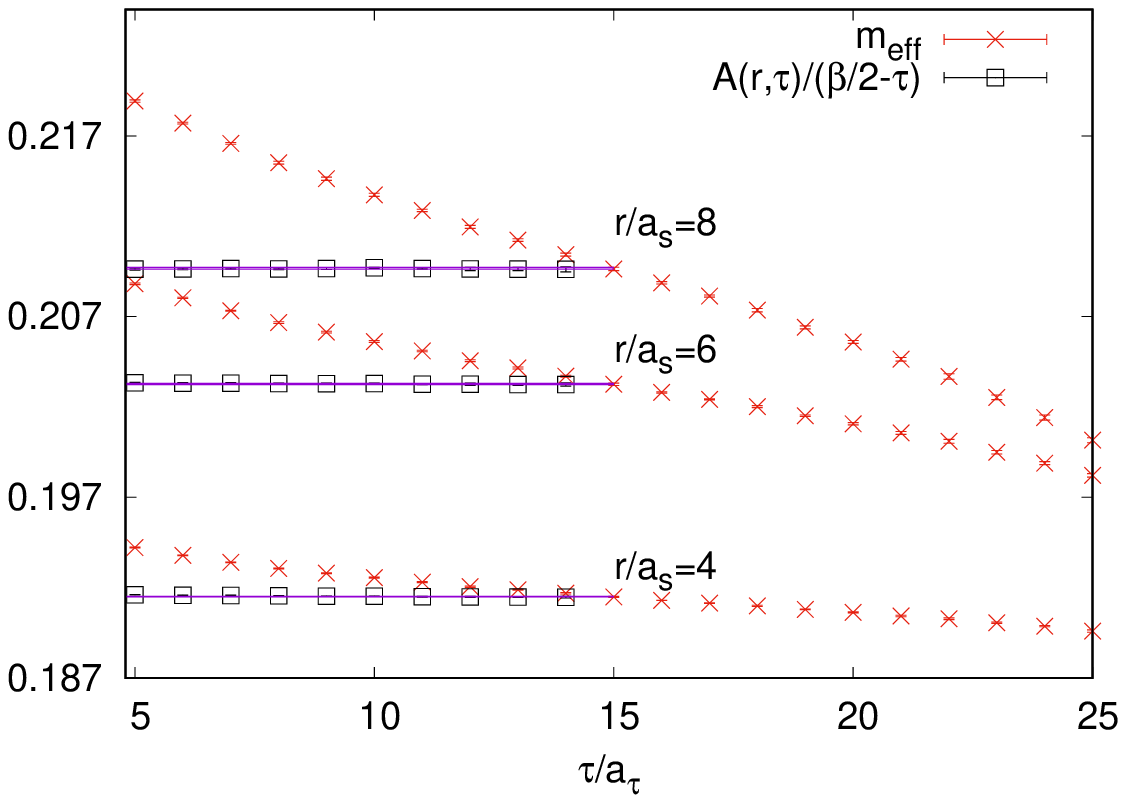}\hspace{1cm}
\includegraphics[width=7cm]{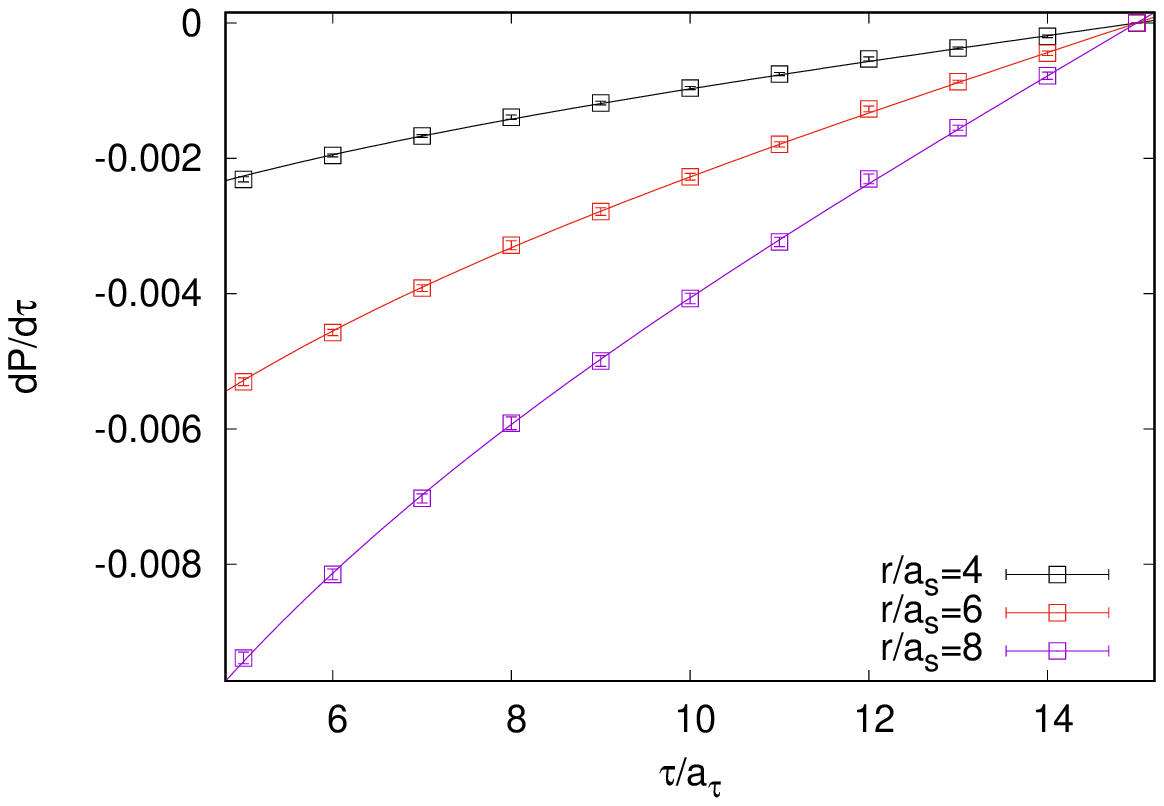}}
\caption{(Left) Comparison of effective mass for $m(r,\tau)$ and
  $A(r,\tau)/(\beta/2-\tau)$ for 1.5 $T_{c}$.
  (Right) $\frac { dP(r,\tau)}{d\tau}$; see the discussion around
  Eq.(\ref{eq.per2}).}
\label{fig.plateau} \end{figure}

Let us first focus on the anti-periodic part $A(r,\tau)=\frac{1}{2}
\log\left(\frac{\langle w_{E}(r,\tau)\rangle}{\langle
  w_{E}(r,\beta-\tau) \rangle}\right)$. From perturbation theory one
expects $A(r_,\tau)= (\frac{\beta}{2}-\tau) \vre$.  In the left panel
of Fig 1 we have plotted the effective mass $m(r,\tau)$ and
$\frac{A(r,\tau)}{({\beta}/{2}-\tau)}$. From the figure it is clear
that non-perturbatively also $A(r,\tau)$ is linear in $\tau$ around
$\beta/2$. For the remaining part, $P(r,\tau)=\frac{1}{2}
\log({\langle w_{E}(r,\tau)\rangle}{\langle
  w_{E}(r,\beta-\tau)\rangle})$, we can again take help from
perturbation theory and write it as
\begin{equation}
  P(r,\tau)=\int_{-\infty}^{\infty} d\omega \; \sigma(\omega) \,
  (e^{-\omega \tau}+e^{-\omega
  (\beta-\tau)} \, + \, \tau \ {\rm independent \ parts}). 
\label{eq.per1}\end{equation}
Now to have a potential $ i \partial_t P(r,it) $ should approach a
constant as $t$ goes to infinity.
\begin{equation}
i\partial_t P(r,it)=\int_{-\infty}^{\infty}(e^{-i \omega t
}-e^{-\omega (\beta-i t)}) \, \omega \, \sigma(\omega) \; d\omega.
\label{eq.per2}\end{equation}
Therefore we need $\sigma(\omega)$ to go like $\sigma(\omega) \sim
\frac{1}{\omega^2}$ as $\omega$ approaches zero. On the right panel of
Fig. \ref{fig.plateau} we have shown a fit of $\partial_\tau P(r,\tau)$
calculated from lattice data with the leading $\frac{1}{\omega^{2}}$
behaviour. We can see that almost the entire range of $\tau$ can be
fitted with this leading singular structure.

One of the $ 1/\omega $ in $1/\omega^2 $ comes from a factor $ 1 +
n_{b}(\omega)$, where $n_{b}(\omega) $ is the Bose distribution
function \cite{laine}. This can be understood from the structure of
the time-ordered correlator \cite{our}.  Therefore we expand
$\sigma(\omega)$ in the following form,
\begin{equation}
  \sigma(\omega)=(1+n_{b}(\omega))\left(\frac{\beta V_{im}}{2\pi\omega}+c_{1}
  \omega + c_{2} \omega^3 + ... \right).
\label{eq.series}\end{equation}
Only the odd terms are present because the even ones do not contribute
to Eq. (\ref{eq.per1}). A very good fit of the data can also be found just by
using two terms in the series. The potential is obtained from the
coefficient of $ 1/\omega $ in the expansion, as only this term contributes in
the long time $t$ limit. Using Eq. (\ref{eq.series}) one would then get
\begin{eqnarray}
  \partial_\tau P(r,\tau) &=& V_{\rm im} \, \cot(\frac{\pi \tau}{\beta}) \; + \;
  c_{1} \, G_{1}(\tau,\beta) \; + \; c_{2} G_{2}(\tau,\beta) \; + \; ...
  \label{eq.imfit} \\
G_{n}(\tau,\beta) &=& \frac{2 (2 n)!}{\beta^{2 n + 1}} \left( \zeta \left( 2
  n+1, \frac{\tau}{\beta} \right) \; - \; \zeta \left( 2 n+1,
  1-\frac{\tau}{\beta} \right) \right). \nonumber
\end{eqnarray}

Our strategy for calculating potential is then obvious: we do a linear
fit of $A(r,\tau)$ near $\beta/2$ to get the real part of the
potential and the imaginary part can be obtained by fitting
Eq. (\ref{eq.imfit}) More details on the analysis can be found in \cite{our}.

\section{Results for singlet channel}
\label{sec.singlet}
We show here the results of the singlet potential obtained from
anisotropic lattices with coupling $\beta=6.64$ and the bare anisotropy
$\xi_{b}=2.55$, which corresponds to a renormalized anisotropy $\xi=3$
and a spatial lattice spacing of 0.048 fm. For results with other lattice
parameters and discussion of cutoff effects see \cite{our}. The spatial
volume was kept fixed at 1.44 fm, and temperatures upto 2 $T_c$ were
explored by varying the temporal extent $N_t$. We used APE
smearing for the spatial links of the Wilson loop. The potential was
calculated with various number of smearing steps, where each smearing step
involved replacing the spatial links $U^i_{\vec{x}}$ by 
${\rm Proj}_{SU(3)} \left(\alpha \, U^i_{\vec{x}} \ +
\ \sum_{j \ne i} \, U^j_{\vec{x}} \, U^i_{\vec{x}+\hat{j} a} \,
U^{j \dagger}_{\vec{x}+\hat{i} a} \right)$, with $\alpha$ = 2.5. 

\begin{figure}
\centerline{\includegraphics[width=7cm]{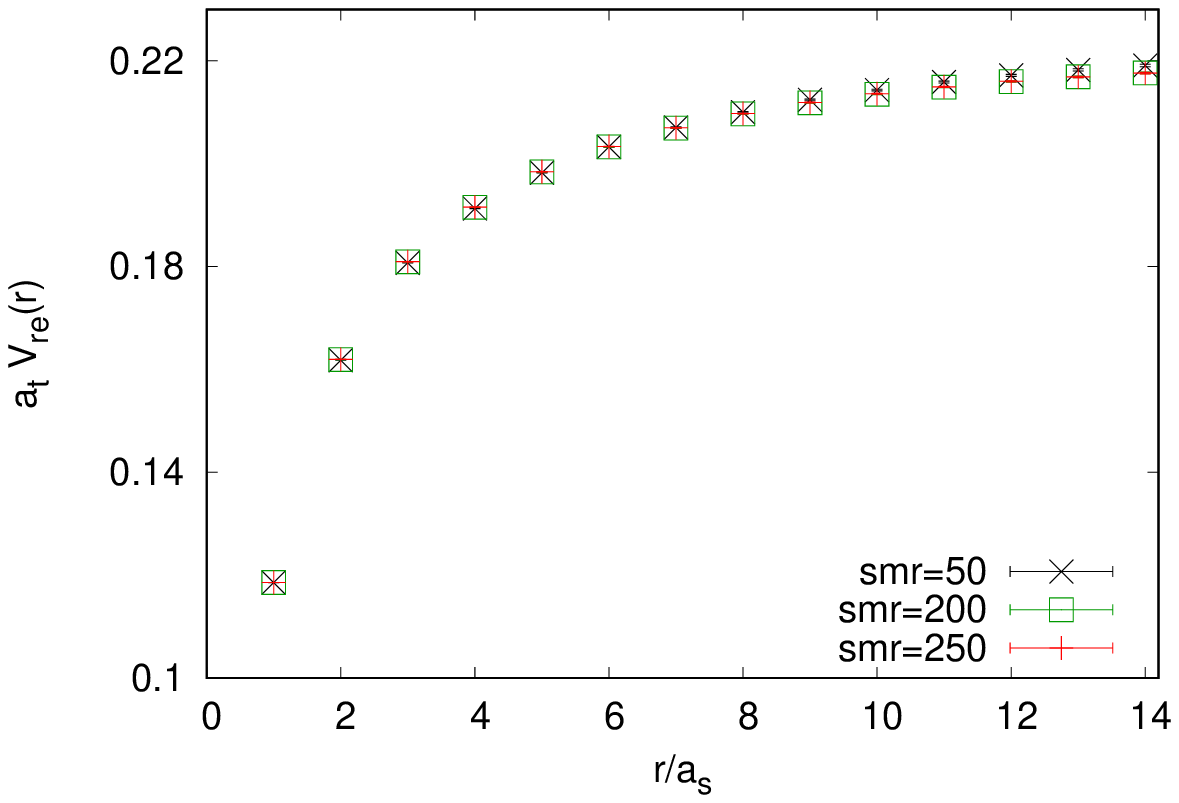}\hspace{1cm}
\includegraphics[width=7cm]{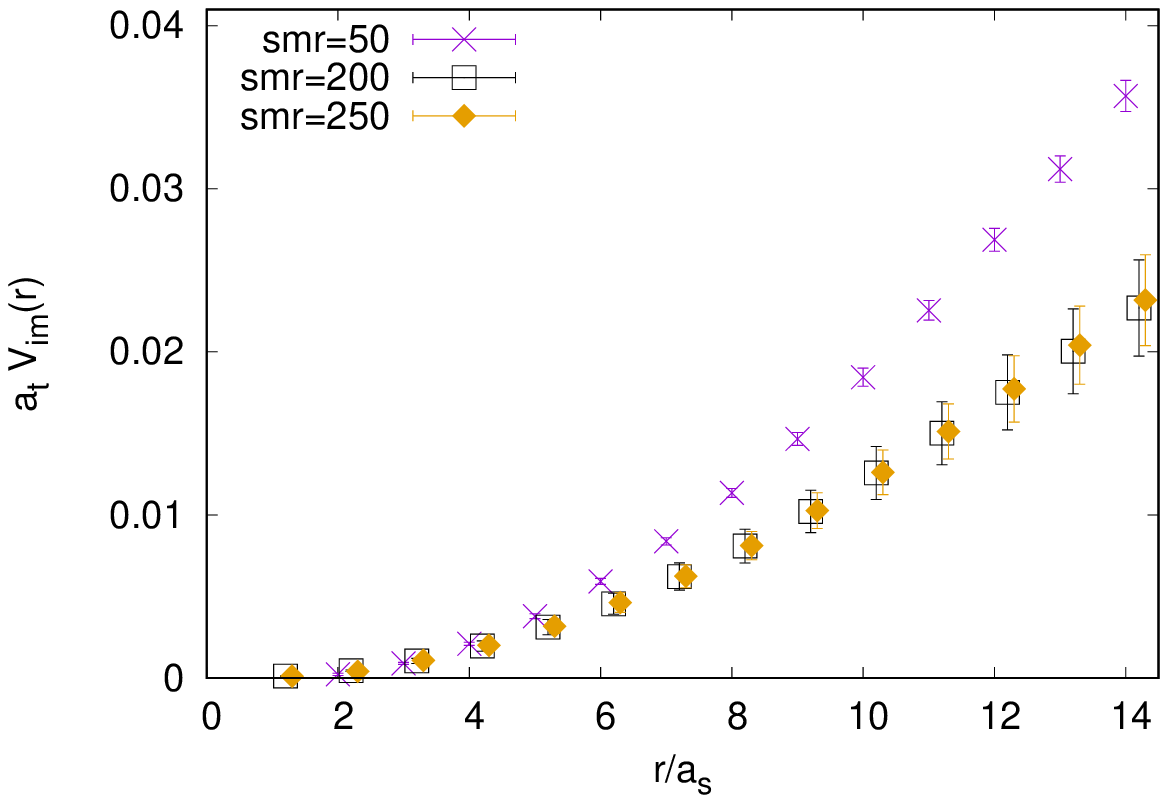}}
\caption{(Left) $\vre$ and (right) $\vim$ at 1.5 $T_c$,
  calculated with different levels of APE smearing.}
\label{fig.smear}
\end{figure}

In Fig. \ref{fig.smear} we have plotted the real and imaginary parts
of the potential as a function of the number of smearing sweeps for
$T=1.5T_c$. From the figure it is clear that the potential has very
minor dependence on the smearing after a certain number of
smearing. Anyway for the calculation of error we have included the
variation with respect to smearing as a systematic error. When quoting
the value of imaginary part the error also includes the variation of
the fit results when we change the number of terms in Eq. (\ref{eq.series}).

\begin{figure}
\centerline{\includegraphics[width=7cm]{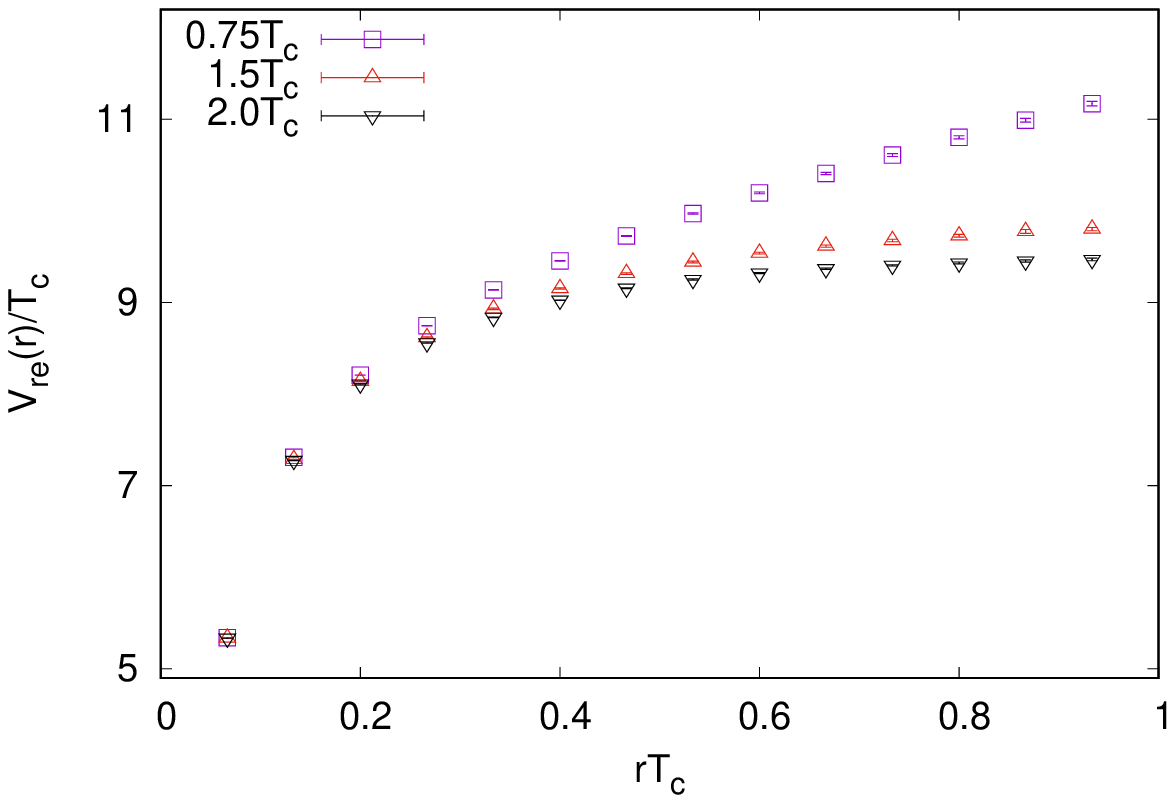}\hspace{1cm}
\includegraphics[width=7cm]{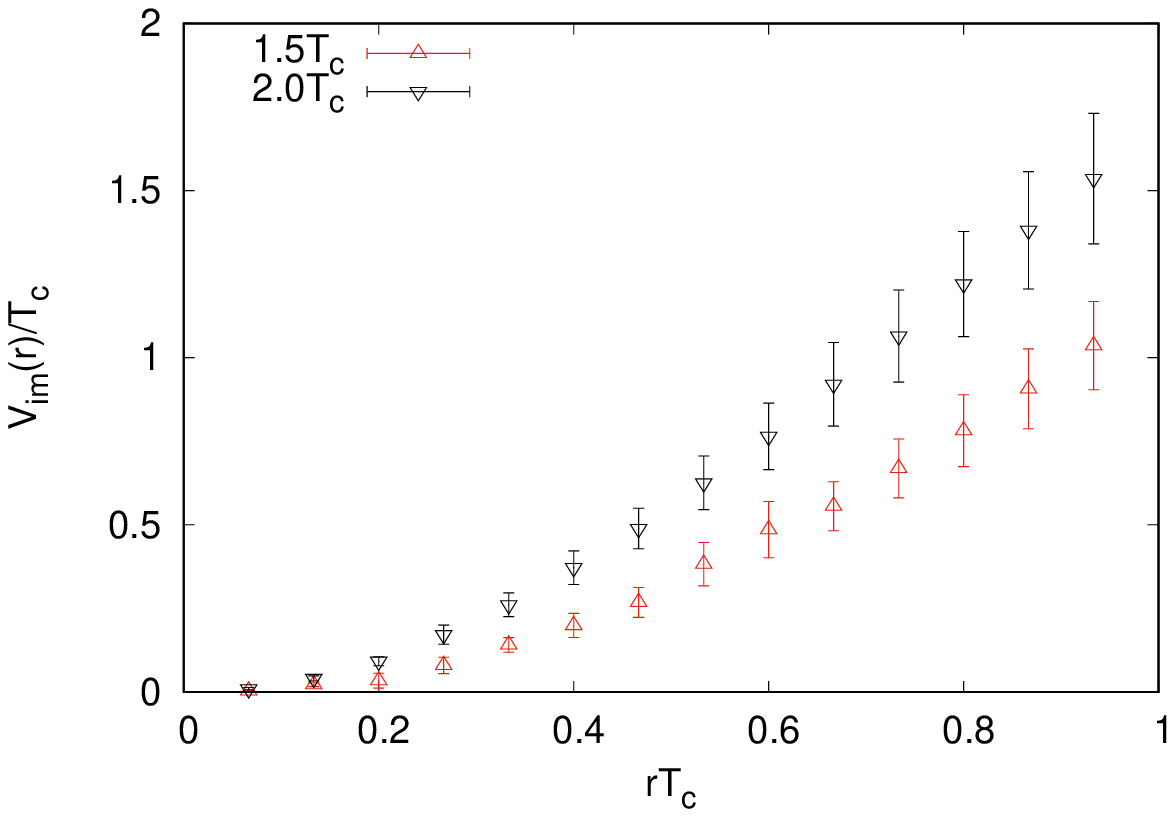}}
\caption{(Left) $\vre$ and (right) $\vim$
  at different temperatures.}
\label{fig.singlet}
\end{figure}

In Fig. \ref{fig.singlet} we show the temperature dependence of the
singlet potential.  $\vre$, shown in the left panel, shows a clear screening
behavior above $T_c$. At a temperature of 0.75 $T_c$ $\vre$ shows the usual
linear string tension behavior at long distances. Above $T_c$ this linear rise
is screened, with the screening increasing with temperature. However,
quantitatively the potential is different from the screened Coulomb form;
addition of a screened string tension term \cite{string} is needed to fit
$\vre$ at these temperatures.

On the right hand side we have plotted $\vim$. The results below $T_c$
are consistent with zero and are not shown in the plot. Above $T_c$,
$\vim$ is very different from the perturbative results of Ref. \cite{laine}.
It is also seen to increase with rise in temperature. More details about the
temperature dependence and parametrization of $\vim$ can be found in
Ref. \cite{our}.

\section{Results for octet channel}
\label{sec.octet}

For phenomenology of quarkonia in QGP, one also needs to understand the
interaction between $Q$ and $\bar{Q}$ when they are in a color octet
configuration. In analogy with the singlet state, we can try to define
a potential using the point-split operator  $\bar{\psi}(\vec x,t)
U(\vec x,\vec x_{0};t) T_{a} U(\vec x_{0},\vec y;t) \psi(\vec y,t)$
\cite{bazavov}. This state, however, is not gauge invariant.

A gauge invariant state with the quark content of the above operator
can be formed by adding a color-adjoint gluonic operator $H_{a}$
\cite{bali}:
\begin{equation}
\mathcal{O}(r=|\vec x - \vec y| ,t; \vec{x}_0) = \bar \psi(\vec x,t)
U(\vec x,\vec x_{0};t)T_{a} H_{a}(\vec x_{0},t) U(\vec x_{0},\vec y;t)
\psi(\vec y,t).
\label{eq.opoctet} \end{equation}
Taking the above source will, in the static limit,
lead to Wilson loop with inserted $T_a H_{a}(\vec{x}_0)$ at both
initial and final time slices. Here we take two operators, $B_a^z$ and
$B_a^x+i B_a^y$, for $H_a$, where the quark and the antiquark are
taken to be separated in the $z$ direction.  These are the hybrid
states with gluonic angular momentum $L=0$ and $L=1$ along $z$
direction for $B_{a}^{z}$ and $B_{a}^{x}+i B_{a}^{y}$
respectively. The results shown in this section are from Wilson loops with
200 steps of smearing for spatial links.

\begin{figure}
\centerline{\includegraphics[width=7cm]{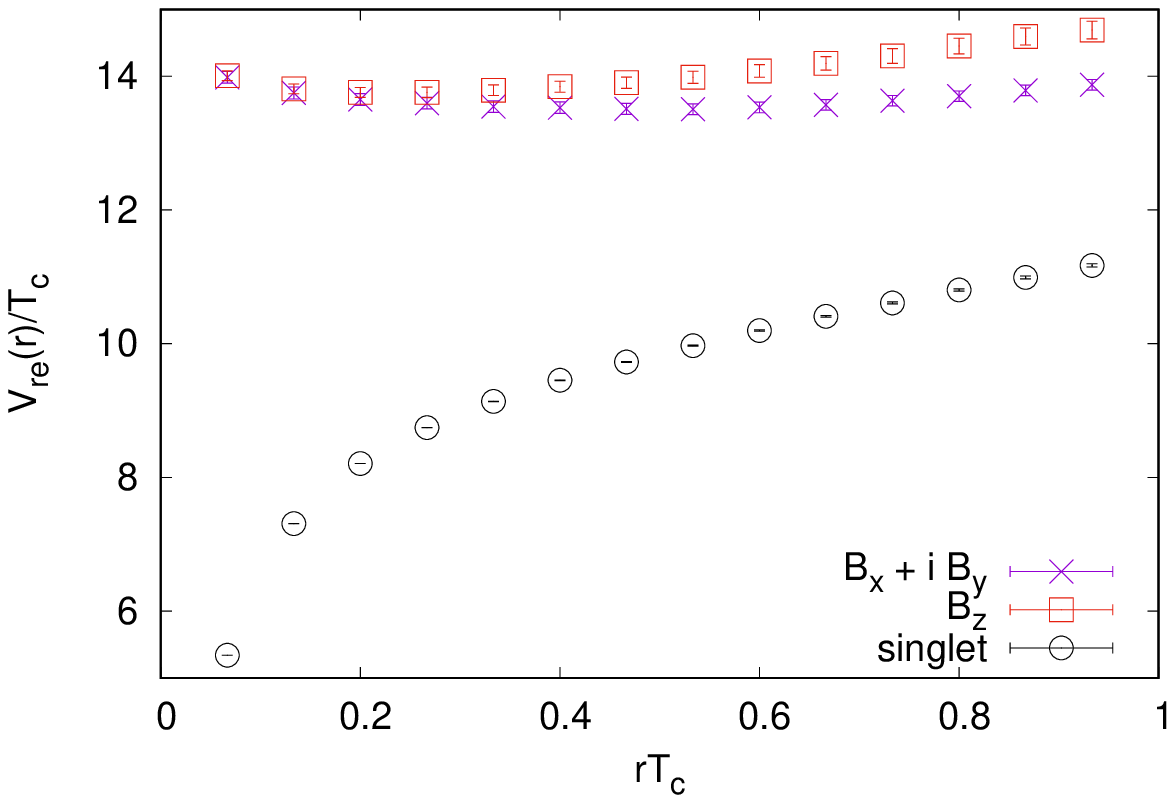}\hspace{1cm}
\includegraphics[width=7cm]{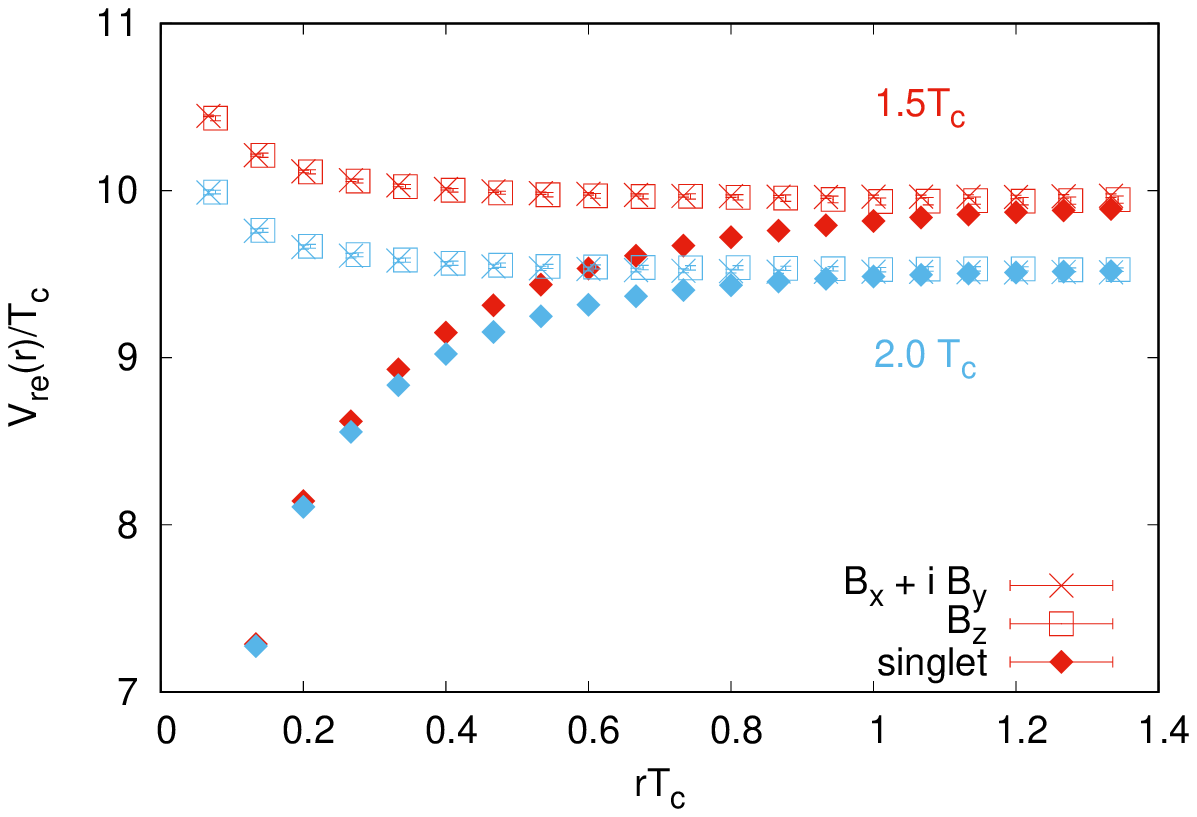}}
\caption{Singlet and ``octet'' potentials between static $Q$ and
  $\bar{Q}$ below (left, at 0.75 $T_c$) and above (right) the
  deconfinement transition temperature.}
\label{fig.octet}
\end{figure}

In the left panel of Fig. \ref{fig.octet} we show the octet potential
at the temperature of 0.75 $T_c$. For comparison, the singlet
potential at this temperature is also shown. As with the singlet
channel, below $T_c$ the potential is largely insensitive to
temperature and closely resembles the zero temperature
potential.  In leading order perturbation theory these hybrid
potentials at short distances only get contribution from octet
channel, and the potential is independent of $H_a$ \cite{bali} at
short distance. However, at long distance the potential depends on
$H_a$ \cite{bali}, leading to different hybrid potentials for the
L = 0 and L = 1 channels.

At finite temperature we have calculated the real part of the
potential for the state $\mathcal{O}(r,t)$ using the method of
Sec. \ref{sec.method}. The plateau structure here is not as good as
that of the singlet, however the real part of the potential can be
obtained by fitting a few points near $\beta/2$. Preliminary results
for the potential are shown in the right panel of
Fig. \ref{fig.octet}. Here we find that the potential is identical for
the two choices of $H_a$ we have used. This suggests that at finite
temperature the effect of $H_a$ gets decoupled from the potential at
all distances, and we can meaningfully talk about the real part of the
effective thermal potential for octet $Q \bar{Q}$. Of course, it would
be good to further check this with other choices of the gluonic
operator.

In the figure we have also shown $\vre$ for the singlet potential, at
the same temperatures. As the figure shows, the octet and singlet
potential approach each other at long distances. At higher
temperatures they approach each other at shorter distances. 
We stress that we did not add any additive
renormalization constant to the octet potential to match with singlet:
as the effect of $H_a$ gets decoupled, the renormalization of the
octet gets fixed once the additive renormalization for the
singlet is fixed.

\begin{figure}
\centerline{\includegraphics[width=7cm]{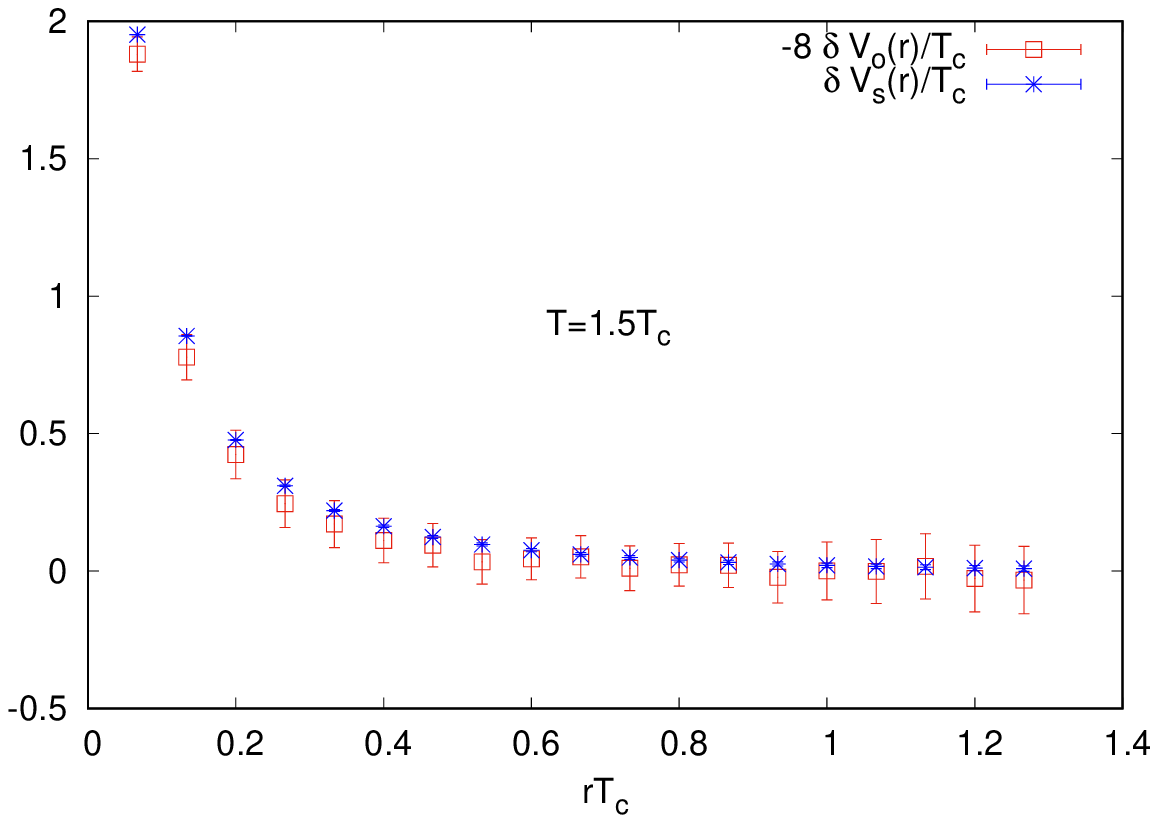}\hspace{1cm}
\includegraphics[width=7cm]{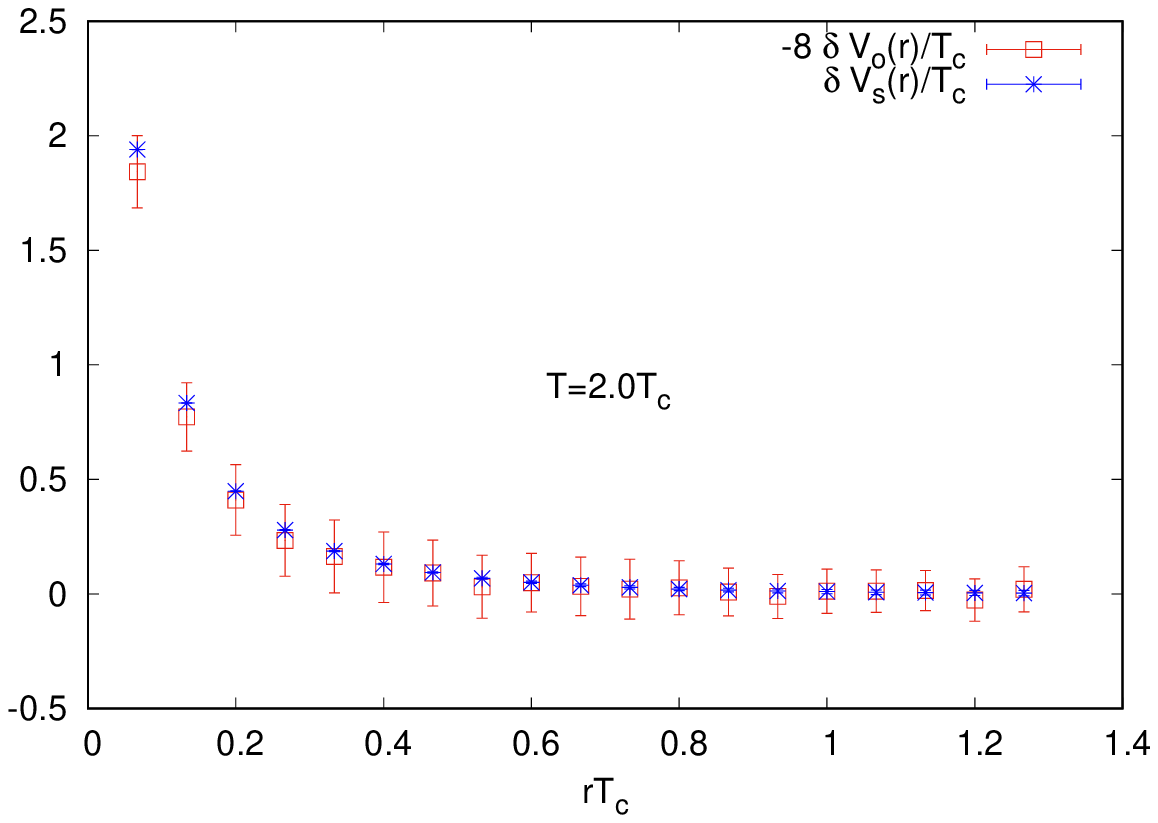}}
\caption{Comparison of octet $\delta V_{o}$ and singlet $\delta
  V_{s}$, at  (left) 1.5 $T_{c}$ and (right) 2.0 $T_{c}$. Here for
  octet, the results for $H_a=B_a^x+i B_a^y$ are shown.}
\label{fig.delta}
\end{figure}

At short distances, the singlet and octet potentials are attractive
and repulsive, respectively, consistent with perturbation theory. To
further investigate conformity with perturbation theory without having
to worry about the additive renormalization constant, we define \cite{owe}
$ \delta V_{o,s}(r)= V_{o,s}(r+a_{s})-V_{o,s}(r)$, where
$o,s$ stand for octet and singlet, respectively.  In leading order of
perturbation theory, $\delta V_{s}(r)=-8 \delta V_{o}(r)$. The
nonperturbative estimates, shown in Fig. \ref{fig.delta}, agree with
this prediction within our errorbars. 

\section{Summary}
Extraction of a thermal potential from the Euclidean time Wilson loop
is difficult and often involves Bayesian analysis. Here we
calculate the complex ``thermal potential'' \cite{laine} using various
properties of Wilson loops, motivated from perturbation theory; the method 
is described in Sec.  \ref{sec.method}. We have discussed results for the
singlet potential in Sec. \ref{sec.singlet}. On crossing $T_c$, the
linear confining part of the real part of the potential is screened,
the screening increasing with increase in temperature. However,
at least upto temperatures of 2 $T_c$ the singlet $\vre$ is different
from the perturbative potential. In the deconfined phase, the
effective potential also acquires an imaginary part. The imaginary
part is very different from the perurbative result, with the potential
not saturating upto the distance scale $r T_c \sim 1.4$, or $r \sim 1$
fm.  We have also studied the effective thermal potential between the
$Q$ and $\bar{Q}$ in an octet configuration. Preliminary results for
the real part of the potential are discussed in
Sec. \ref{sec.octet}. Our results indicate that, unlike at zero
temperature, the thermal potential between hybrid states is not
sensitive to the gluonic structure in $\mathcal{O}(r)$,
Eq. (\ref{eq.opoctet}). While at short distances the octet potential
is repulsive, at long distances it approaches the singlet potential.
Qualitatively the potential is similar to the free energy for the
octet state \cite{olaf}.

We acknowledge support of the Department of Atomic Energy, 
Government of India, under project no. 12-R\&D-TFR-5.02-0200. The
computations reported here were performed on the clusters of the
Department of Theoretical Physics, TIFR. We would like to thank Ajay
Salve and Kapil Ghadiali for technical support.


\begin{thebibliography}{99}
\bibitem{laine}
M. Laine, O. Philipsen, P. Romatschke \& M. Tassler, JHEP 0703 (2007) 054.
\bibitem{akamatsu}
Y. Akamatsu, Phys. Rev. D 87 (2013) 045016; Phys. Rev. D 91 (2015) 056002.
\bibitem{hatsuda}
A. Rothkopf, T. Hatsuda \& S  Sasaki, Phys. Rev. Lett. 108 (2012) 162001.
\bibitem{rothkopf}
Y. Burnier, O. Kaczmarek \& Alexander Rothkopf, Phys. Rev. Lett.
  114 (2015) 082001 (2015).
\bibitem{peter}
P. Petreczky \& J. weber, Nuclear Physics A 967 (2017) 592.
\bibitem{our}
D. Bala \& S. Datta, arXiv:1909.10548 [hep-lat]. 
\bibitem{string}
Y. Burnier \& A. Rothkopf, Phys. Lett. B 753 (2016) 232
\bibitem{bazavov}
A Bazavov \& P Petreczky 2013 J. Phys: Conf. Ser. 432, 012003.
\bibitem{bali}
G.S. Bali \& A. Pineda, Phys. Rev. D 69 (2004) 094001.
\bibitem{owe}
O. Philipsen, Phys. Lett. B 535 (2002) 138.  
\bibitem{olaf}
  F. Zantow, O. Kaczmarek, F. Karsch \& P. Petreczky, {\em
    Proceedings, 5th International Conference on Strong and
    Electroweak Matter}, World scientific, 2003 (hep-lat/0301015). 
\end{thebibliography}
\end{document}